# Anomalous phase transitions in one-dimensional organo-metal halide perovskite nanorods grown inside porous silicon nanotube templates

N. Arad-Vosk,[1] N. Rozenfeld,[1] R. Gonzalez-Rodriguez,[2] J. L. Coffer[2] and A. Sa'ar[1,*]

[1] *Racah Institute of Physics and the Harvey M. Kruger family center for Nanoscience and Nanotechnology, the Hebrew University of Jerusalem, Jerusalem 91904, Israel*

[2] *Department of Chemistry, Texas Christian University, Fort Worth, Texas, 76129 USA*

## Abstract

One-dimensional organo-metal halide Perovskite ($CH_3NH_3PbI_3$) nanorods whose diameter and length are dictated by the inner size of porous silicon nanotube templates have been grown, characterized and compared to bulk perovskites in the form of microwires. We have observed a structural phase transition for bulk perovskites, where the crystal structure changes from tetragonal to orthorhombic at about 150K, as opposed to small diameter one-dimensional perovskite nanorods, of the order of 30-70 nm in diameter, where the phase transition is inhibited and the dominant phase remains tetragonal. Two major experimental techniques, infrared absorption spectroscopy and photoluminescence, were utilized to probe the temperature dependence of the perovskite phases over the 4-300K temperature range. Yet, different characteristics of the phase transition were measured by the two spectroscopic methods and explained by the presence of small, tetragonal inclusions embedded in the orthorhombic phase. The inhibition of the phase transition is attributed to the large surface area of these one-dimensional perovskite nanorods, which gives rise to a large stress that, in turn, prevents the formation of the orthorhombic phase. The absence of phase transition enables the measurement of the tetragonal bandgap energy down to low temperatures.

---

* Email: saar.amir@mail.huji.ac.il

# I. INTRODUCTION

Hybrid organo-metal halide perovskites have attracted much attention over recent years, mainly due to the high conversion efficiency (of sun light to electricity, exceeding 20%[1–4]) achieved for these perovskite-based solar cells. Originally, perovskites of the generic formula, $CH_3NH_3PbX_3$ (X=I, Br, Cl), have been utilized to replace conventional sensitizers in dye-sensitized solar cells (DSSCs) where they were used for absorbing the sun light spectrum.[5–8] However, soon after it has been realized that perovskites are characterized by good electronic and optoelectronic properties,[9–16] that can be exploited for advancing thin-film and low-dimensional[17–21] versions of solar cells, as well as for developing perovskite-based light emitting diodes (LEDs)[22–24] and lasers.[25–30]

The rapid progress in the field of perovskites has also raised certain questions concerning fundamental properties of these materials, particularly about phenomena associated with low-dimensional effects and size-dependent properties of perovskite nanostructures. $CH_3NH_3PbX_3$ perovskites are composed of organic and inorganic constituents that form a complex hybrid having properties that can be tuned between its organic and inorganic counterparts. For example, the free carriers mobility and diffusion length in these perovskites are similar to those of inorganic semiconductors[31–38] while the ease of preparation[39] resembles that of organic complexes.[40,41] Here we focus on another unique property of perovskites associated with the crystal structure and structural phase transitions taking place upon heating and cooling of the material.[42] Recent investigation of bulk perovskites by variety of experimental techniques (such as neutron powder diffraction,[43] single crystal x-ray diffraction,[44] calorimetric and infrared spectroscopy[45]) revealed structural phase transitions, from cubic to tetragonal phase above room-temperature and from tetragonal to orthorhombic phase at low-temperatures (~ 160K).[42,45–50] In addition, it has been found recently that the phase transition temperature exhibits size-dependence for thin-films of perovskites[51] (which can be considered as quasi-2D film of perovskites once the thickness of the 2D film drops below 100 nm). Understanding basic properties of these nanostructures, particularly the photo-physics associated with structural changes due to phase transitions, is the main objective of this report.

A unique method to produce and control the size and the dimensionality of $CH_3NH_3PbI_3$ perovskites has been reported by us recently.[52] In brief, by careful selection of the inner diameter of porous Si nanotube (PSiNT) templates, we have succeeded to control the diameter of one-dimensional (1D) perovskite nanorods grown inside these PSiNTs. Here, we investigate how the size and the dimensionality of these hybrid organo-metal halide perovskites affect structural characteristics, particularly the phase transition and the photo-physics[50] of perovskite nanostructures.

## II. EXPERIMENTAL

The method to produce one-dimensional $CH_3NH_3PbI_3$ perovskite nanorods having variable diameters has been described in details in reference [52]. In brief, the fabrication process that is schematically illustrated in Fig. 1, involves initial production of ZnO nanowire sacrificial templates followed by CVD deposition of Si (using $SiH_4$ as a gas source at ~ 530$^o$C) on top of the ZnO nanowires template, followed by template removal by $NH_3$/HCl etch.[53] As a result, a matrix of PSiNTs is formed where the inner diameter of the nanotubes is determined by the diameter of the original ZnO nanowires. Three different inner diameters (175, 70 and 30 nm) were chosen for the current study. Two types of substrates, either F-doped tin oxide (FTO) that is transparent to visible light or Si substrate that is transparent across the infrared range of the electromagnetic spectrum, were used. Next, $CH_3NH_3PbI_3$ perovskites were loaded into the PSiNTs by immersing the PSiNTs array in 1:1 solution of $CH_3NH_3I$ and $PbI_2$ (206 mM in DMF) for 2 hours at 60°C. Excess reactants were removed by spin coating, followed by thermal annealing at 95°C for 30 minutes. This process was repeated three times to ensure a more uniform loading of perovskites into the PSiNTs and minimizing the amount of perovskites outside the nanotubes. A reference 3D array of bulky perovskites, in the form of microwires, has also been prepared using a threefold more concentrated solution of perovskite precursor,[52] yielding microwires having average diameter of about 8 μm. Fig. 2 shows high-resolution images of unloaded (Fig. 2a) and loaded (Fig. 2b) PSiNTs obtained using transmission electron microscope (TEM; FEI-Technai F20 G2 operated at 200 kV). The porous nature of the nanotube walls can clearly be appreciated from these images as well as the formation of small perovskite nanorods whose diameter is determined by the inner diameter of the nanotubes. More images of similar perovskite nanorods can be found in reference [52].

For temperature-dependent photoluminescence (PL) spectroscopy, the samples were excited by Ar$^+$ ion laser operating at 488 nm, while the PL signal was dispersed with a 1/4-m monochromator and detected by a photomultiplier tube (PMT). During PL experiments the samples were kept under vacuum in a continuous-flow liquid helium optical cryostat that allows temperature control over the 4.5K to 300K range. The same optical cryostat has been used for temperature-dependent infrared (IR) absorption spectroscopy, where the cryostat was placed inside the sample compartment of Bruker's Vertex V70 Fourier Transform Infrared (FTIR) spectrometer (Bruker Optik GmbH, Ettlingen, Germany), equipped with deuterated triglycine sulphate (DTGS) detector. Transmission ($T$) measurements were performed at normal incidence, and the absorbance ($A$) was taken to be, $A$=-ln$(T)$. In both experiments (PL and IR absorption spectroscopies), the samples were first cooled down to 4.5K and then heated up gradually back to room temperature.

## III. RESULTS

Fig. 3 shows three-dimensional (3D) plots of the PL versus photon energy and temperature for, (a) bulk perovskites in the form of microwires and, (b-d) perovskite nanorods (loaded into PSiNTs) having average diameter of 175, 70 and 30 nm respectively. Clearly, two groups of emission peaks can be identified from the spectra (marked by vertical dotted lines in Fig. 3), particularly for the 3D bulk perovskites shown in Fig. 3(a) and the 175 nm (in diameter) perovskite nanorods of Fig. 3(b). The first group is related to a low-energy emission line at about 1.57-1.63 eV, which is relatively weak at room-temperature and becomes stronger with the decreasing temperature. This low-energy emission line can be observed for all four samples of bulk perovskites and perovskite nanorods. The second group of high-energy emission lines appears at higher photon energies, in the range of 1.66-1.67 eV, and can be observed essentially for the 3D bulk perovskites and the 175 nm nanorods at temperatures below 140K, but not for the 70 and the 30 nm nanorods. A more profound means to analyze these emission lines is by comparing their relative lineshape and strength at different temperatures (around the phase transition temperature of about 120-140K); see Fig. 4. Here, the normalized PL intensity is plotted at different temperatures where the relative strength and the lineshape of the PL for all four samples can be compared at each temperature. As will be discussed in more details hereafter, we argue that the PL spectra can be divided into two sets of samples; the first set includes the bulk perovskites and the 175 nm nanorods that present both PL emission groups (at low-energies and at high-energies) while the second set of 70 nm and 30 nm nanorods showing only the low-energy emission group at all temperatures. Furthermore, the low-energy emission group has been attributed to PL from the tetragonal phase of perovskites that is characterized by a smaller bandgap energy relative to the orthorhombic phase to which, the high energy emission group is assigned.[42,45,46–48] Hence, both bulk perovskites and 175 nm nanorods samples show the expected structural phase transition from tetragonal phase to orthorhombic phase at about 140K, where above 140K only the tetragonal phase appears (*i.e.*, the low-energy emission line) while both the orthorhombic and the tetragonal phases can be observed at temperatures below 140K. Similar phase-transition characteristics of 3D (bulk) perovskites has been observed by others[46–50] and supported by theoretical calculations.[54,55] This implies that, as far as the PL spectroscopy is concerned, both the microwires and the 175 nm nanorods behave as 3D perovskites.  On the other hand, only the low-energy emission group has been observed for the 70 nm and the 30 nm nanorods over all temperatures, indicating that the tetragonal phase dominates throughout the entire temperature range and the phase-transition is inhibited in these essentially 1D perovskite nanorods.

In order to quantify dynamical characteristics of the phase transition, we have used the following procedure. The areas below the low-energy (tetragonal) and the high-energy (orthorhombic) PL emission lines have been estimated by fitting the experimental data to superposition of Gaussian-shape functions according to the number of peaks observed for each spectrum shown in Fig. 4 (for each sample and at each temperature). The relative area of the low-energy group, (*i.e.*, the area of the orthorhombic line normalized to the total area below the entire PL spectrum) versus temperature is presented in Fig. 5.  Let us point out that the relative area can be associated with the fractional volume of the orthorhombic phase, however, one should remember that the data has not been corrected for spectral variation of the absorption cross section and the PL emission efficiency. The results presented in Fig. 5 clearly indicate a phase transition for the 3D bulk perovskites and the 175 nm nanorods while no phase transition takes place for the 70 nm and 30 nm nanorods. The phase transition is not abrupt as expect from a first-order phase transition, in contrast to other experiments and theory.[45,56]   This point will further be elaborated in the next section.

Next, kinetics of the phase transition has been studied by plotting the peak energy, as extracted from the Gaussian fitting procedure, versus temperature for each sample. The results, plotted in Fig. 6, can be divided again into two sets; the first set consists of bulk perovskites and 175 nm nanorods [Fig. 6(a-b)] where the tetragonal (low-energy) peak energies (red symbols) show non-monotonic behavior, starting with a red-shift of the peak energy with the decreasing temperature down to the phase-transition temperature (120-140K), followed by abrupt change into a blue-shift during the phase-transition, and remaining approximately constant when the temperature is further goes down.  In contrast, the orthorhombic (high-energy) peak energies (blue symbols and lines in Fig.6) present a monotonic red-shift with the decreasing temperature below the phase-transition temperature (these high-energy peaks disappear from the PL spectra above the phase transition temperature). On the other hand, the second set of 70 and 30 nm nanorods [Fig. 6(c-d)] presents a single (low-energy) peak energy having quite similar characteristics of a monotonic red-shift with the decreasing temperature until the phase-transition temperature and remains approximately constant below that temperature; see Fig. 6(c-d).

Following our interpretation of the 1D perovskite nanorods <u>not</u> undergoing a structural phase transition, one may conclude that the energy variations shown in Fig. 6(c-d) represent the energy-bandgap variations of the tetragonal phase throughout the entire temperature range. Hence, following O'Donnell *et al*,[57] we have fitted the data of Fig. 6(c-d) to the following expression:

$$E_g(T) = E_g(0) - S\langle\hbar\omega\rangle[\coth(\langle\hbar\omega\rangle/2kT) - 1] \qquad (1)$$

Where $E_g(0)$ - is the bandgap energy at zero temperature, $S$ - is a dimensionless coupling constant and $\langle\hbar\omega\rangle$ is the average phonon energy. The value of these parameters for the tetragonal phase, extracted from the best fit into eq. 1, are summarized in table I and represented by the solid lines in Fig. 6(c-d). Furthermore, the high-energy peaks of Fig. 6(a-b), which are related to the orthorhombic phase and therefore, limited to temperatures below the phase-transition, can also be fitted to the same expression (eq. 1). The blue solid lines in these figures represent the best fits into eq. 1 for the orthorhombic bandgap energy while the extracted fitting parameters for this phase are also summarized in table I.

In another set of experiments, which is quite sensitive to structural changes, the infrared (IR) absorption spectra of similar perovskite structures were measured over the 4-300K temperature range. For these experiments, all perovskite samples were grown on top of Si substrates, which are transparent at the infrared wavelengths. Fig. 7 presents IR absorption spectra over the 850-3600 cm$^{-1}$ wavenumber range and at different temperatures around the phase transition.[+] The spectra consist of essentially three groups of vibrational modes, which can be identified, analyzed and compared to the literature, mainly for the bulk (3D) perovskites at low-temperatures; see Fig. 7(a). For example, at 4K [Navy-blue line in Fig. 7(a)] the first group over the 900-920 cm$^{-1}$ wavenumbers consists of two sharp peaks at 909 and 919 cm$^{-1}$, which can be assigned to the $CH_3NH_3$ rocking modes of the orthorhombic phase,[45,58] while a second group of relatively broader peaks at 1450 and 1457 cm$^{-1}$ have been assigned to the $CH_3$ and the $NH_3$ bending modes of that phase respectively.[58] Finally, a third group of 4 peaks at 3027, 3078, 3121 and 3172 cm$^{-1}$ were assigned to the C-H and the N-H stretching modes, again, all related to the orthorhombic phase.[58]

While these vibrational modes of the orthorhombic phase are relatively narrow and well defined, heating up the samples up to about the phase transition temperature and above, gives rise to a significant broadening and intensity change of the peaks. In particular, the low-frequency rocking mode at ~ 909 cm$^{-1}$ becomes much broader during the phase transition where the higher frequency rocking mode at ~ 919 cm$^{-1}$ becomes much weaker and eventually disappears from the spectra above 160K, particularly as the tail of the much stronger (and broader) peak at 909 cm$^{-1}$ screens this vibration. Moving to the second group of vibrations, one can see that the overlapping CH$_3$ and NH$_3$ bending modes almost vanish from the spectra above 160K and another vibrational mode at ~ 1468 cm$^{-1}$ emerges at 160K and above [Fig. 7(a)].[59] Finally, the four narrow peaks of the third group that characterize bulk perovskites at low-temperatures, merge into a much broader IR absorption line at 140K with the spectrally narrow features that characterize the orthorhombic phase, effectively disappear above that temperature. Quite different results have been measured for all 1D perovskites including the 175 nm (Fig.7b) and the 30 nm (Fig. 7c) nanorods. Here, the broad IR absorption features that characterize the tetragonal phase above the phase transition temperature do not vanish with the decreasing temperature, even well below the phase transition temperature. While for the 175 nm nanorods one can still observe a weak contribution of the narrower vibrations that characterize the orthorhombic phase, these narrow features practically disappear from the spectra of the 30 nm nanorods, even at 4K.

As the results of the IR absorption spectroscopy seem to qualitatively resemble those of the PL experiments, it seems natural to apply a similar analysis of the phase transition dynamics to the IR absorption experiment. However, spectral overlap between some of the vibrations (below and above the phase transition temperature) substantially complicates the analysis of the area below the absorption peaks. Hence, we have chosen to analyze the dynamics using the following characteristic vibrations, represented by the (black) vertical-dotted lines in Fig.7. (a) The 919 cm$^{-1}$ rocking mode that can be observed at low-temperatures (for bulk perovskites), but disappear above the phase transition temperature; this rocking mode was chosen to characterize the orthorhombic phase. (b) The 1468 cm$^{-1}$ bending mode that persists for bulk perovskites above the phase transition only, was taken to be a characteristic of the tetragonal phase. The results, related again to the relative area of the orthorhombic phase, are shown in Fig. 8. Let us point out that quite similar results can be obtained for other vibrational modes (such as the 3027 cm$^{-1}$ and 3078 cm$^{-1}$ stretching modes) but with much larger errors due to overlap of the peaks and their lower strength. Fig. 8 qualitatively shows a similar phenomenon to that observed in the PL experiment, *e.g.*, a phase transition from the tetragonal to the orthorhombic phase taking place again for bulk (3D) perovskites and almost completely inhibited for 1D nanorods. Yet, the characteristics of the phase transition, *e.g.*, the phase transition temperature, width and particularly the type of transition seem to be quite different when comparing the two experiments (PL and IR absorption spectroscopy)!   This is fairly unusual and ambiguous result as the same type of perovskite samples used for both experiments. This topic will further be discussed and explained in the next section.

## IV. DISCUSSION

At first, let us point out that the emergence of a structural phase transition for 3D bulk hybrid perovskites,  from a tetragonal phase above the phase transition temperature to an orthorhombic phase below that temperature, have experimentally been reported and theoretically been analyzed by numerous groups.[42,46–50,54,55,60–64] Furthermore, there are certain publications describing direct evidences to the existence of two phases at low temperatures,[48,65–67] which are also supported by our own low-temperature TEM electron diffraction experiments reported in the supplementary information.  Hence, the most significant finding of this work is the inhibition of the phase transition in 1D perovskite nanorods having diameter below ~ 100-150 nm. This conclusion is supported by both PL and IR absorption experiments, thus indicating that the tetragonal phase of 1D perovskite nanorods is more stable than the orthorhombic phase down to low temperatures.   Let us point out that similar experiments using thin films of hybrid perovskites having thickness smaller than ~ 150 nm, which might be considered as the 2D counterparts of the 1D perovskite nanorods studied here, revealed a related phenomenon of a decreasing phase transition temperature with the decreasing thickness of the 2D thin films.[51]  Hence, summarizing these findings for hybrid organo-metal perovskites one may conclude that the dimensionality of the perovskite nanostructures gives rise to lowering phase transition temperature in 2D and complete inhibition of the phase transition in 1D. To follow the origin of this scarce dimensional phenomenon in hybrid perovskites, let us point out that (apart of quantum size effects[17,18]) the dimensionality of small semiconductor nanostructures gives rise to enlargement of the surface-to-volume (STV) ratio, thus amplifying the role of surface energy (relative to energies and phenomena associated with the bulk crystal). It has been demonstrated that, if $d$ - is a typical size that characterizes the surface of a given dimensional nanostructure and $R$ – being a characteristic length-scale of its volume than, the STV ratio in 2D is given by, $STV=d/R$, while in 1D the ratio is twice larger, $STV=2(d/R)$.[68] Yet, one still need to explore the mechanism responsible to a larger surface energy associated with the orthorhombic phase relative to that connected with the tetragonal phase.

Such a mechanism has been proposed by Ong *et al.*,[69] who have suggested that the origin of the structural phase transition (in 3D) should be related with the volume of a unit cell of the crystal, particularly the ratio of out-of-plane (*c*) to in-plane (*a*) lattice constants, (*c/a*). Below a critical value of this ratio, $c/a \cong 1.45$, the orthorhombic phase is the more stable state of bulk perovskites, while above this ratio the tetragonal phase is the more stable phase with its bulk (or volume) minimum energy laying below that of the orthorhombic phase. In low-dimensional nanostructures one should also take into account the surface energy, for example, in our case of perovskite nanorods embedded in PSiNTs, the surface energy associated with the interface between perovskites and silicon. Yet, the smaller volume of a unit cell of Si (~ 160 Å$^3$) compared to that of perovskites (~ 1000 Å$^3$), give rise to a compressive stress on the perovskite unit cell that, in turn, preferred the tetragonal phase with the larger *c/a* ratio. This mechanism seems to be consistent with our findings of phase transition inhibition in 1D and the decreasing phase transition temperature in 2D, as well as providing a mechanical origin to the structural phase transition in 3D.[69] Notice also a recent report about the possibility of phase transition inhibition in thin films of perovskites stabilized with PMMA,[70] a result that might also be attributed to interfacial stress.

A second unusual and ambiguous result that has been observed in our experiments is related to the different characteristics of the phase transition as obtained from PL experiments (Fig. 5) and from IR absorption experiments (Fig. 8). In fact, there are two major observations that seem to be inconsistent when comparing the two experiments:

(1) The results shown in Fig. 8 (based on IR absorption spectroscopy) can be considered as a clear indication of a 1st order phase-transition taking place around 150-160K for 3D bulk perovskites, in a very good agreement with other observations based on IR spectroscopy[15] and theoretical calculations.[54,55,71] Also, the first derivative of the orthorhombic relative area for bulk perovskites supports this claim of a 1st order phase transition having a width of less than 6K; see, supporting information. Yet, the results obtained via IR absorption spectroscopy seem to disagree with those obtained via PL experiments, particularly for 3D bulk perovskites; see Fig. 5. In this figure, the phase transition extends from about ~ 150K down to ~ 70K with a gradual change of the orthorhombic area. Let us emphasize that quite similar, temperature-dependence PL-based experimental results for bulk perovskites, were reported by other groups.[46–48,61,66,72] Yet, neither discussion of the nature and the type of phase transition nor explanation of the discrepancy between the two observations, has been reported so far.

(2) In addition, considering the complementary area of the tetragonal phase, $\Lambda_T$ (defined as, $\Lambda_T = 1 - \Lambda_O$; where $\Lambda_T$ and $\Lambda_O$ are the relative areas of the tetragonal and the orthorhombic phase respectively), then the tetragonal area for the bulk perovskites) goes down to about 70% (of its original value) below the phase transition temperature, as opposed to the IR absorption experiment where the tetragonal area goes down to ~ 0 below that temperature.

These disputing results can be explained as follows. It has been argued that fairly small inclusions of tetragonal nanoparticles exist inside the (dominant) orthorhombic phase of the bulk (3D) perovskites, below the phase transition temperature.[48,67] Furthermore, the smaller bandgap energy of the tetragonal phase (relative to the orthorhombic phase with, $\Delta E_g \cong$ 60-80 meV; see Fig. 9) compared to $kT$ below the phase transition temperature ($kT <$ 14 meV), favors the transfer of excited free carriers (which are absorbed mainly in the orthorhombic phase) into the lower bandgap tetragonal phase followed by a radiative recombination (of electrons and holes) from that phase.[48,65,66] This mechanism significantly amplifies and broadens the PL emission from the tetragonal inclusions, despite that the fractional volume of this phase is practically very small below the phase transition temperature.

Finally, let us point out that the above model of tetragonal inclusions within the orthorhombic phase, can nicely explain the kinetic results obtained for the low-energy PL lines, see Figs. 6(a-b); [e.g., the red lines in these figures]. Above the phase-transition temperature these PL lines of bulk perovskites (and of 175 nm nanorods), which are related to PL emission from the tetragonal phase, show indeed a monotonic red-shift with the decreasing temperature, similar to that observed for 1D nanorods [Fig. 6(c-d)]. However, once approaching the phase transition (at ~ 150K), the low-energy lines should be attributed to the residual small inclusions of the tetragonal phase within the (dominant) orthorhombic phase.[48] As the orthorhombic phase is characterized by a smaller unit cell volume relative to the tetragonal phase,[54] one should expect induced (compressive) stress on the tetragonal inclusions to appear. As a result, the induced stress gives rise to a blue-shift of the bandgap energy as qualitatively observed for 3D perovskites; see, Figs. 6(a-b).

## V CONCLUSIONS

In summary, we have investigated the physical properties of 1D hybrid $CH_3NH_3PbI_3$ perovskite nanorods, grown inside PSiNT templets, which allow good control over the shape and the diameter of these perovskite nanorods. We have studied the influence of the nanorods dimensionality on the structural phase transition, from a tetragonal phase into an orthorhombic phase, that characterizes 3D (bulk) perovskites. We found that the structural phase transition is inhibited for small diameter perovskite nanorods having a diameter smaller than ~ 100-150 nm, thus indicating a better stability of perovskite nanorods. This result might be significant for novel applications involving the use of perovskites such as photovoltaics, lasers and optoelectronics.

Two spectroscopic methods, temperature-dependent PL and IR absorption, were utilized for this research. However, different characteristics of the phase transition, particularly the type of transition and its width, were observed for each technique. We have explained the disputing results by the presence of small tetragonal inclusions within the orthorhombic phase, below the phase transition temperature, which cause the PL-based characteristics of the phase transition to differ from those obtained via IR absorption spectroscopy. Finally, we have assigned the inhibition of the phase transition to the large surface area in these 1D nanorods that gives rise to a significant contribution of surface energy associated with stress, which favors the tetragonal phase over the orthorhombic phase. The inhibition of the phase transition in 1D nanorods enables us to measure the tetragonal energy bandgap down to low temperatures.

# References

[1] W.S. Yang, J.H. Noh, N.J. Jeon, Y.C. Kim, S. Ryu, J. Seo, and S. Il Seok, Science **348**, 1234 (2015).

[2] H.P. Zhou, Q. Chen, G. Li, S. Luo, T.B. Song, H.S. Duan, Z.R. Hong, J.B. You, Y.S. Liu, and Y. Yang, Science **345**, 542 (2014).

[3] H. Kim, J. Lee, N. Yantara, P.P. Boix, S. a. Kulkarni, S. Mhaisalkar, M. Grätzel, and N. Park, Nano Lett. **13**, 2412 (2013).

[4] T. Salim, S. Sun, Y. Abe, A. Krishna, A.C. Grimsdale, and Y.M. Lam, J. Mater. Chem. A Mater. Energy Sustain. **3**, 8943 (2015).

[5] A. Kojima, K. Teshima, Y. Shirai, and T. Miyasaka, J. Am. Chem. Soc. **131**, 6050 (2009).

[6] J.-H. Im, C.-R. Lee, J.-W. Lee, S.-W. Park, and N.-G. Park, Nanoscale **3**, 4088 (2011).

[7] S.P. Singh and P. Nagarjuna, Dalton Trans. **43**, 5247 (2014).

[8] M.A. Green, A. Ho-Baillie, and H.J. Snaith, Nat. Photonics **8**, 506 (2014).

[9] M.M. Lee, J. Teuscher, T. Miyasaka, T.N. Murakami, and H.J. Snaith, Science **338**, 643 (2012).

[10] N.J. Jeon, J.H. Noh, W.S. Yang, Y.C. Kim, S. Ryu, J. Seo, and S. Il Seok, Nature **517**, 476 (2015).

[11] T.C. Sum and N. Mathews, Energy Environ. Sci. **7**, 2518 (2014).

[12] J. Burschka, N. Pellet, S.-J. Moon, R. Humphry-Baker, P. Gao, M.K. Nazeeruddin, and M. Grätzel, Nature **499**, 316 (2013).

[13] D. Liu and T.L. Kelly, Nat. Photonics **8**, 133 (2013).

[14] C. Wehrenfennig, M. Liu, H.J. Snaith, M.B. Johnston, and L.M. Herz, J. Phys. Chem. Lett. **5**, 1300 (2014).

[15] M. Grätzel, Nat. Mater. **13**, 838 (2014).

[16] Q. Lin, A. Armin, R.C.R. Nagiri, P.L. Burn, and P. Meredith, Nat. Photonics **9**, 106 (2015).

[17] J.R. Klein, O. Flender, M. Scholz, K. Oum, and T. Lenzer, Phys. Chem. Chem. Phys. **18**, 10800 (2016).

[18] M.E. Kamminga, H.-H. Fang, M.R. Filip, F. Giustino, J. Baas, G.R. Blake, M.A. Loi, and T.T.M. Palstra, Chem. Mater. **28**, 4554 (2016).

[19] J. Even, L. Pedesseau, and C. Katan, ChemPhysChem **15**, 3733 (2014).

[20] D.N. Dirin, L. Protesescu, D. Trummer, I. V Kochetygov, S. Yakunin, F. Krumeich, N.P. Stadie, and M. V. Kovalenko, Nano Lett. **xxx**, xxx (2016).

[21] M.J. Ashley, M.N. O 'brien, K.R. Hedderick, J.A. Mason, M.B. Ross, and C.A. Mirkin, **138**, 10096 (2016).

[22] S.D. Stranks and H.J. Snaith, Nat. Nanotechnol. **10**, 391 (2015).

[23] Z.-K. Tan, R.S. Moghaddam, M.L. Lai, P. Docampo, R. Higler, F. Deschler, M. Price, A. Sadhanala, L.M. Pazos, D. Credgington, F. Hanusch, T. Bein, H.J. Snaith, and R.H. Friend, Nat. Nanotechnol. **9**, 687 (2014).

[24] J. Pan, L.N. Quan, Y. Zhao, W. Peng, B. Murali, S.P. Sarmah, M. Yuan, L. Sinatra, N. Alyami, J. Liu, E. Yassitepe, Z. Yang, O. Voznyy, R. Comin, M.N. Hedhili, O.F. Mohammed, Z.H. Lu, D.H. Kim, E.H. Sargent, and O.M. Bakr, Adv. Mater. **Accepted**, 1 (2016).

[25] H. Zhu, Y. Fu, F. Meng, X. Wu, Z. Gong, Q. Ding, M. V Gustafsson, M.T. Trinh, S. Jin, and X.-Y. Zhu, Nat. Mater. **14**, 636 (2015).

[26] F. Deschler, M. Price, S. Pathak, L.E. Klintberg, D.D. Jarausch, R. Higler, S. Hüttner, T. Leijtens, S.D. Stranks, H.J. Snaith, M. Atatüre, R.T. Phillips, and R.H. Friend, J. Phys. Chem. Lett. **5**, 1421 (2014).

[27] G. Xing, N. Mathews, S.S. Lim, N. Yantara, X. Liu, D. Sabba, M. Grätzel, S. Mhaisalkar, and T.C. Sum, Nat. Mater. **13**, 476 (2014).

[28] Q. Zhang, S.T. Ha, X. Liu, T.C. Sum, and Q. Xiong, Nano Lett. **14**, 5995 (2014).

[29] T.S. Kao, Y.H. Chou, C.H. Chou, F.C. Chen, and T.C. Lu, Appl. Phys. Lett. **105**, 231108 (2014).

[30] R. Dhanker, A.N. Brigeman, A. V. Larsen, R.J. Stewart, J.B. Asbury, and N.C. Giebink, Appl. Phys. Lett. **105**, 151112 (2014).

[31] Y. Zhao, A.M. Nardes, and K. Zhu, J. Phys. Chem. Lett. **5**, 490 (2014).

[32] Q. Dong, Y. Fang, Y. Shao, P. Mulligan, J. Qiu, L. Cao, and J. Huang, Science **347**, 967 (2015).

[33] G. Xing, N. Mathews, S.S. Lim, Y.M. Lam, S. Mhaisalkar, and T.C. Sum, **6960**, 498 (2013).

[34] D. Shi, V. Adinolfi, R. Comin, M. Yuan, E. Alarousu, A. Buin, Y. Chen, S. Hoogland, A. Rothenberger, K. Katsiev, Y. Losovyj, X. Zhang, P.A. Dowben, O.F. Mohammed, E.H. Sargent, and O.M. Bakr, Science **347**, 519 (2015).

[35] S.D. Stranks, G.E. Eperon, G. Grancini, C. Menelaou, M.J.P. Alcocer, T. Leijtens, L.M. Herz, A. Petrozza, and H.J. Snaith, Science **342**, 341 (2013).

[36] C.S. Ponseca, T.J. Savenije, M. Abdellah, K. Zheng, A. Yartsev, T. Pascher, T. Harlang, P. Chabera, T. Pullerits, A. Stepanov, J.P. Wolf, and V. Sundstrom, J. Am. Chem. Soc. **136**, 5189 (2014).

[37] W. Nie, H. Tsai, R. Asadpour, J.-C. Blancon, A.J. Neukirch, G. Gupta, J.J. Crochet, M. Chhowalla, S. Tretiak, M.A. Alam, H.-L. Wang, and A.D. Mohite, Science **347**, 522 (2015).

[38] M.A. Loi and J.C. Hummelen, Nat Mater **12**, 1087 (2013).

[39] S. Brittman, G.W.P. Adhyaksa, and E.C. Garnett, MRS Commun. **5**, 7 (2015).

[40] J. Meyer, S. Hamwi, M. Kroger, W. Kowalsky, T. Riedl, and A. Kahn, Adv. Mater. **24**, 5408 (2012).

[41] C. Wang, H. Dong, W. Hu, Y. Liu, and D. Zhu, Chem. Rev. **112**, 2208 (2012).

[42] C.C. Stoumpos, C.D. Malliakas, and M.G. Kanatzidis, Inorg. Chem. **52**, 9019 (2013).

[43] M.T. Weller, O.J. Weber, P.F. Henry, A.M. Di Pumpo, and T.C. Hansen, Chem. Commun. **51**, 4180 (2015).

[44] T. Baikie, Y. Fang, J.M. Kadro, M. Schreyer, F. Wei, S.G. Mhaisalkar, M. Graetzel, T.J. White, M. Gratzel, and T.J. White, J. Mater. Chem. A **1**, 5628 (2013).

[45] N. Onoda-yamamuro, T. Matsuo, and H. Suga, J. Phys. Chem. Solids **51**, 1383 (1990).

[46] R.L. Milot, G.E. Eperon, H.J. Snaith, M.B. Johnston, and L.M. Herz, Adv. Funct. Mater. **25**, 6218 (2015).

[47] K. Wu, A. Bera, C. Ma, Y. Du, Y. Yang, L. Li, and T. Wu, Phys. Chem. Chem. Phys. **16**, 22476 (2014).

[48] C. Wehrenfennig, M. Liu, H.J. Snaith, M.B. Johnston, and L.M. Herz, APL Mater. **2**, 081513 (2014).

[49] A.M. Soufiani, F. Huang, P. Reece, R. Sheng, A. Ho-Baillie, and M. a. Green, Appl. Phys. Lett. **107**, 231902 (2015).

[50] H.H. Fang, R. Raissa, M. Abdu-Aguye, S. Adjokatse, G.R. Blake, J. Even, and M.A. Loi, Adv. Funct. Mater. **25**, 2378 (2015).

[51] D. Li, G. Wang, H.-C. Cheng, C.-Y. Chen, H. Wu, Y. Liu, Y. Huang, and X. Duan, Nat. Commun. **7**, 11330 (2016).

[52] R. Gonzalez-Rodriguez, N. Arad-Vosk, N. Rozenfeld, A. Sa'ar, and J.L. Coffer, Small **12**, 4477 (2016).

[53] X. Huang, R. Gonzalez-Rodriguez, R. Rich, Z. Gryczynski, and J.L. Coffer, Chem. Commun. **49**, 5760 (2013).

[54] T. Oku, in *Sol. Cells - New Approaches Rev.*, edited by Leonid A. Kosyachenko (InTech, 2015), pp. 77–101.

[55] W. Geng, L. Zhang, Y.-N. Zhang, W.-M. Lau, and L.-M. Liu, J. Phys. Chem. C **118**, 19565 (2014).

[56] A. Yangui, M. Sy, L. Li, Y. Abid, P. Naumov, and K. Boukheddaden, Sci. Rep. **5**, 16634 (2015).

[57] K.P. O'Donnell and X. Chen, Appl. Phys. Lett. **58**, 2924 (1991).

[58] M.A. Perez Osorio, R.L. Milot, M.R. Filip, J.B. Patel, L.M. Herz, M.B. Johnston, and F. Giustino, J. Phys. Chem. C **119**, 25703 (2015).

[59] T. Glaser, C. Müller, M. Sendner, C. Krekeler, O.E. Semonin, T.D. Hull, O. Yaffe, J.S. Owen, W. Kowalsky, A. Pucci, and R. Lovrinčić, J. Phys. Chem. Lett. **6**, 2913 (2015).

[60] F. Brivio, J.M. Frost, J.M. Skelton, A.J. Jackson, O.J. Weber, M.T. Weller, A.R. Goni, A.M. a Leguy, P.R.F. Barnes, and A. Walsh, Phys. Rev. B **92**, 144308 (2015).

[61] S. Singh, C. Li, F. Panzer, K.L. Narasimhan, A. Graeser, T.P. Gujar, A. Köhler, M. Thelakkat, S. Huettner, and D. Kabra, Phys. Chem. Lett. **7**, 3014 (2016).

[62] J. Even, L. Pedesseau, and C. Katan, J. Phys. Chem. C **118**, 11566 (2014).

[63] V. D'Innocenzo, G. Grancini, M.J.P. Alcocer, A.R.S. Kandada, S.D. Stranks, M.M. Lee, G. Lanzani, H.J. Snaith, and A. Petrozza, Nat. Commun. **5**, 3586 (2014).

[64] D. Täuber, A. Dobrovolsky, R. Camacho, and I.G. Scheblykin, Nano Lett. **16**, 5087 (2016).

[65] L.Q. Phuong, Y. Yamada, M. Nagai, N. Maruyama, A. Wakamiya, and Y. Kanemitsu, J. Phys. Chem. Lett. **7**, 2316 (2016).

[66] W. Kong, Z. Ye, Z. Qi, B. Zhang, M. Wang, A. Rahimi-Iman, and H. Wu, Phys. Chem. Chem. Phys. **17**, 16405 (2015).

[67] K. Galkowski, A. Mitioglu, A. Surrente, Z. Yang, D.K. Maude, P. Kossacki, G.E. Eperon, J.T.-W. Wang, H.J. Snaith, P. Plochocka, and R.J. Nicholas, arXiv:1606.03234v1 (2016).

[68] A. Saar, J. Nanophotonics **3**, 032501 (2009).

[69] K.P. Ong, T.W. Goh, Q. Xu, and A. Huan, J. Phys. Chem. Lett. **6**, 681 (2015).

[70] T. Dittrich, C. Awino, P. Prajongtat, B. Rech, and M.C. Lux-Steiner, J. Phys. Chem. C **119**, 1 (2016).

[71] X. Zhu, H. Su, R.A. Marcus, and M.E. Michel-Beyerle, J. Phys. Chem. Lett. **5**, 3061 (2014).

[72] Y. Yamada, T. Nakamura, M. Endo, A. Wakamiya, and Y. Kanemitsu, IEEE J. Photovoltaics **5**, 401 (2015).

**Table I:** Summary of the fitting parameters to equation (1) for the various PL peak energies (shown in Fig. 6).

| Sample | Phase | $E_g(0)$ (eV) | $S$ | $\langle \hbar \omega \rangle$ (eV) |
|---|---|---|---|---|
| Bulk perovskites | O | 1.66±0.01 | -1.0±0.2 | 0.009±0.003 |
| 175nm nanorods | O | 1.65±0.01 | -1.4±0.2 | - |
| 70nm nanorods | T | 1.58±0.01 | -2.3±0.3 | 0.037±0.005 |
| 30nm nanorods | T | 1.59±0.01 | -4.4±0.8 | 0.059±0.005 |

O - Orthorhombic
T – Tetragonal

## Figure Captions

**Figure 1:** Schematic view of the perovskite's fabrication stages. Each scheme is composed of a top view of a single object and a side view of several objects on top of the substrate. (a) Growth of ZnO nanowires as a sacrificial template. (b) Chemical vapor deposition (CVD) of Si on top of the ZnO nanowires. (c) Template removal by etching with $NH_3$/HCl gas mixture to create empty nanotubes. (d) Annealing of the empty nanotubes to create PSiNTs. (e) PSiNTs immersed in in 1:1 solution of $CH_3NH_3I$ and $PbI_2$ (f) Spin coating followed by thermal annealing to create PSiNTs loaded with perovskites. Steps (e-f) are repeated three times to ensure uniform loading of the perovskites. (g) perovskite loaded PSiNTs

**Figure 2:** TEM images showing, (a) empty PSiNT and, (b) perovskite loaded PSiNT.

**Figure 3:** 3D plots of the temperature-dependence PL spectra versus photon energy for, (a) bulk perovskites (microwires), (b-d) 175 nm, 70 nm and 30 nm perovskite nanorods respectively.

**Figure 4:** Comparison of the normalized PL spectra of the bulk (black line) and the 1D perovskite nanorods (blue, green and red lines are related to nanorods of 175, 70 and 30 nm in diameter respectively) at different temperatures across the phase-transition temperature (of about 120-140K).

**Figure 5:** The relative area of the low-energy PL line (associated with the orthorhombic phase), normalized to the total area of both the low-energy and the high-energy (associated with the tetragonal phase) PL lines, for the bulk (black sybmols and line), 175 nm nanorods (blue), 70 nm nanorods (green) and 30 nm nanorods (red). All data are related to results of the temperature-dependent PL spectroscopy shown in figures (3-4).

**Figure 6:** Temperature dependence of the PL peak energies associated with the tetragonal phase (low-energy PL peaks; red symbols and lines) and the orthorhombic phase (high-energy PL peaks; blue

symbols and lines), for: (a) bulk perovskites, (b) 175 nm nanorods, (c) 70 nm nanorods and, (d) 30 nm nanorods. Notice that the orthorhombic peaks energies (blue symbols and lines) can be resolved for bulk and 175 nm perovskite nanorods and at temperatures below 140K only.  The solid lines represent the best fit of the experimental data to equation (1) with the fitting parameters summarized in table I.

**Figure 7:** IR absorption spectra of (a) bulk perovskites (microwires), (b) 175 nm nanorods, (c) 70 nm nanorods and, (d) 30 nm nanorods, at various temperatures from 4K up to 180K.  The black vertical dotted lines represent the two vibrational modes (at 919 cm$^{-1}$ and 1468 cm$^{-1}$) chosen to characterize the orthorhombic and the tetragonal phases respectively.

**Figure 8:** The relative area of the orthorhombic phase (the 919 cm$^{-1}$ rocking mode), normalized to the total area of the orthorhombic plus the tetragonal (1468 cm$^{-1}$ bending mode) phases, for bulk perovskites (black symbols and line), 175 nanorods (blue symbols and line) and, 30 nm nanorods (red symbols and line). All data are related to results of the temperature-dependent IR absorption spectroscopy shown in figure (8).

**Figure 9:** Schematic bandgap-energy drawings of the orthorhombic phase ($E_g^O$) and the small tetragonal inclusions having smaller bandgap energy ($E_g^T < E_g^O$). As a result, charge carriers that are excited at the orthorhombic phase via photon absorption, are trapped in the tetragonal inclusions and radiatively recombine from that phase. The values of the bandgap energies of the two phases can be found in table I.

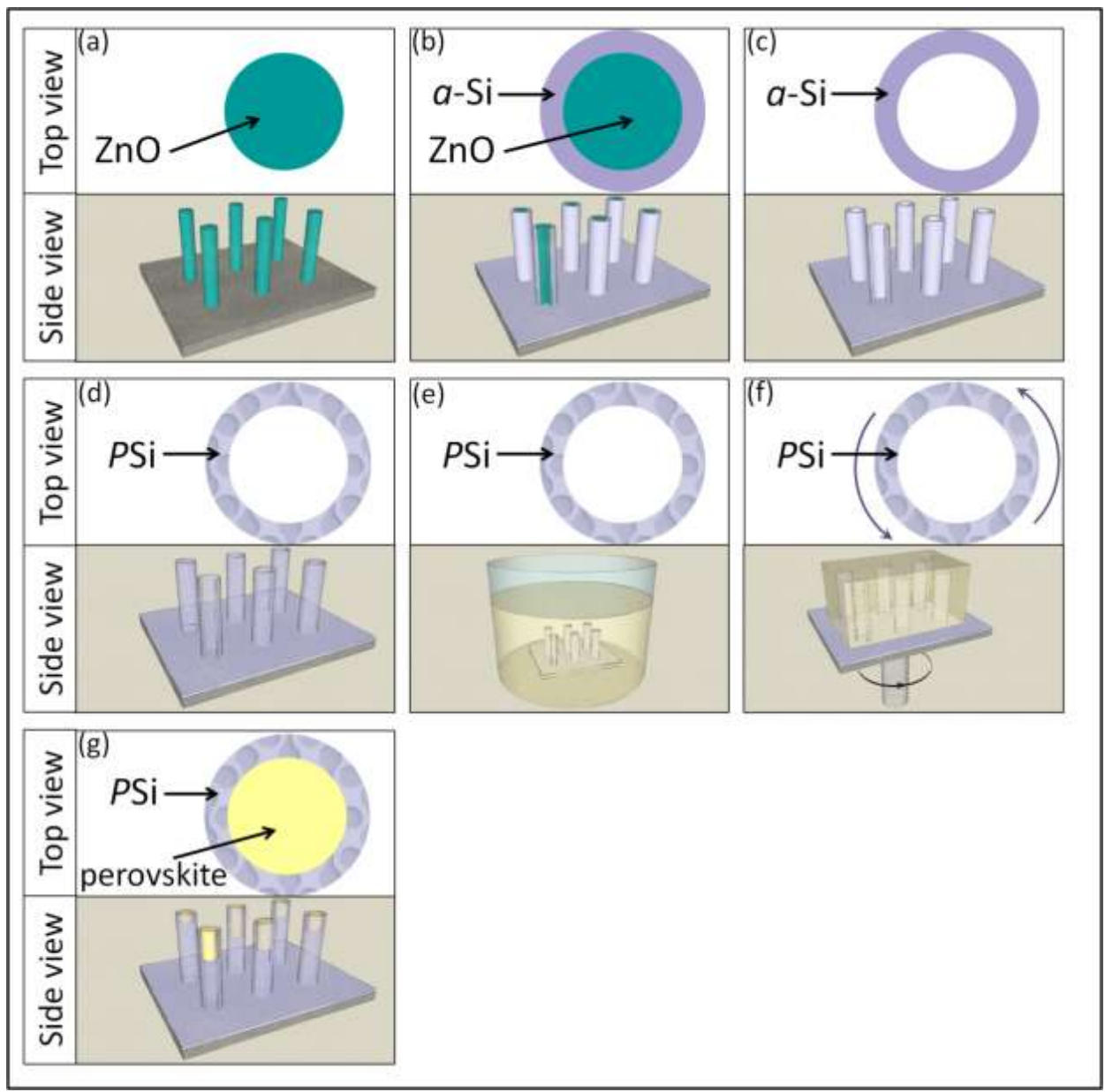

**Figure 1**

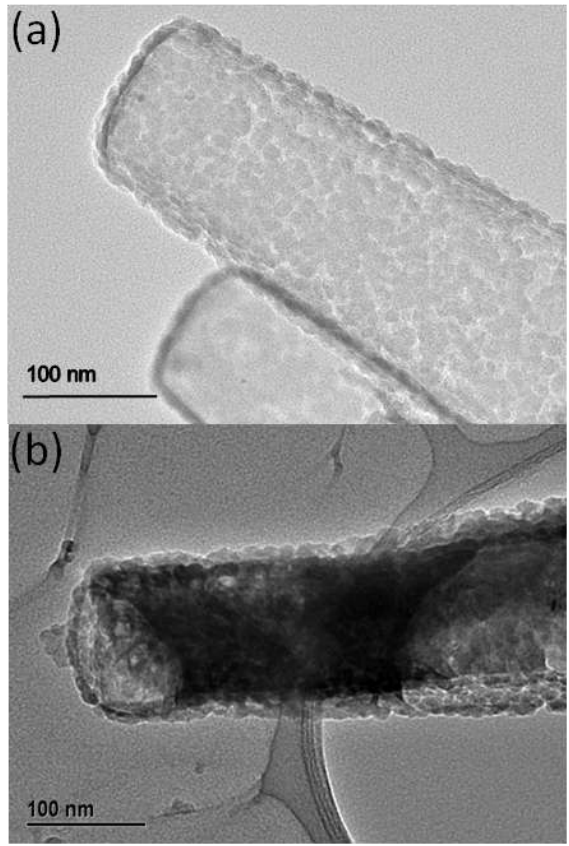

**Figure 2**

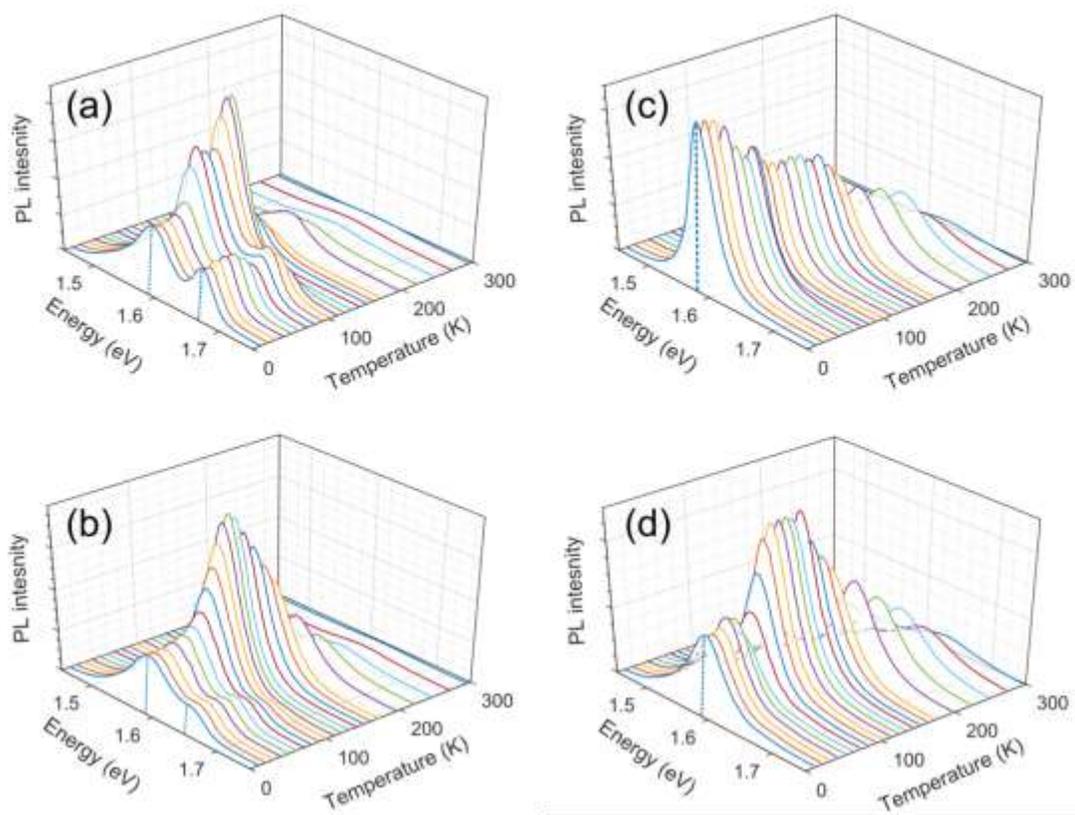

**Figure 3**

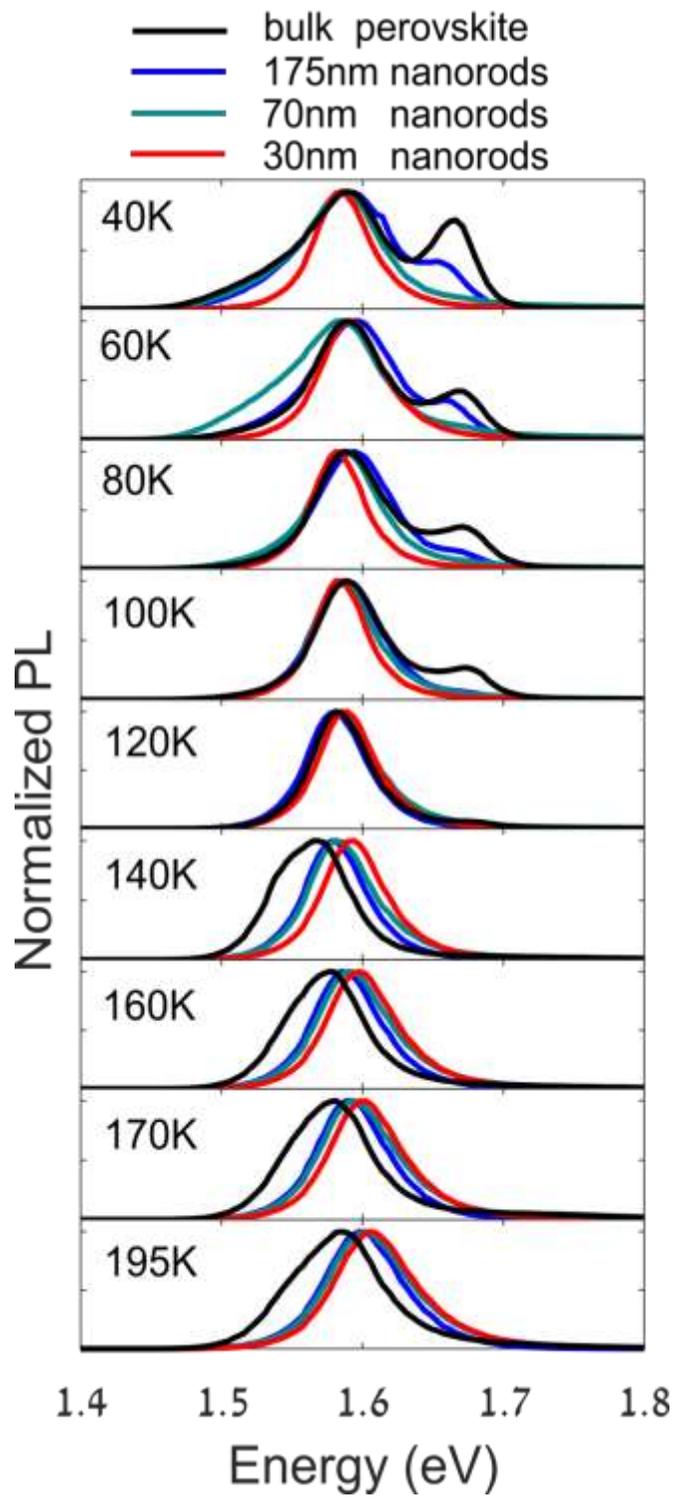

**Figure 4**

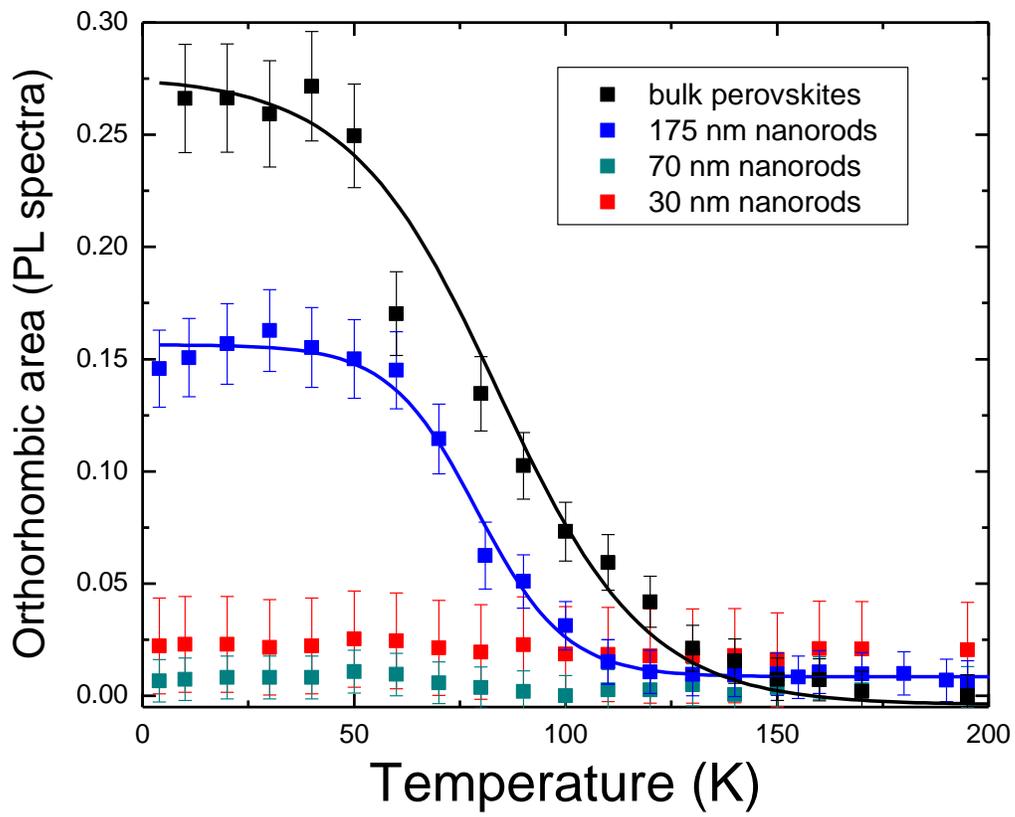

**Figure 5**

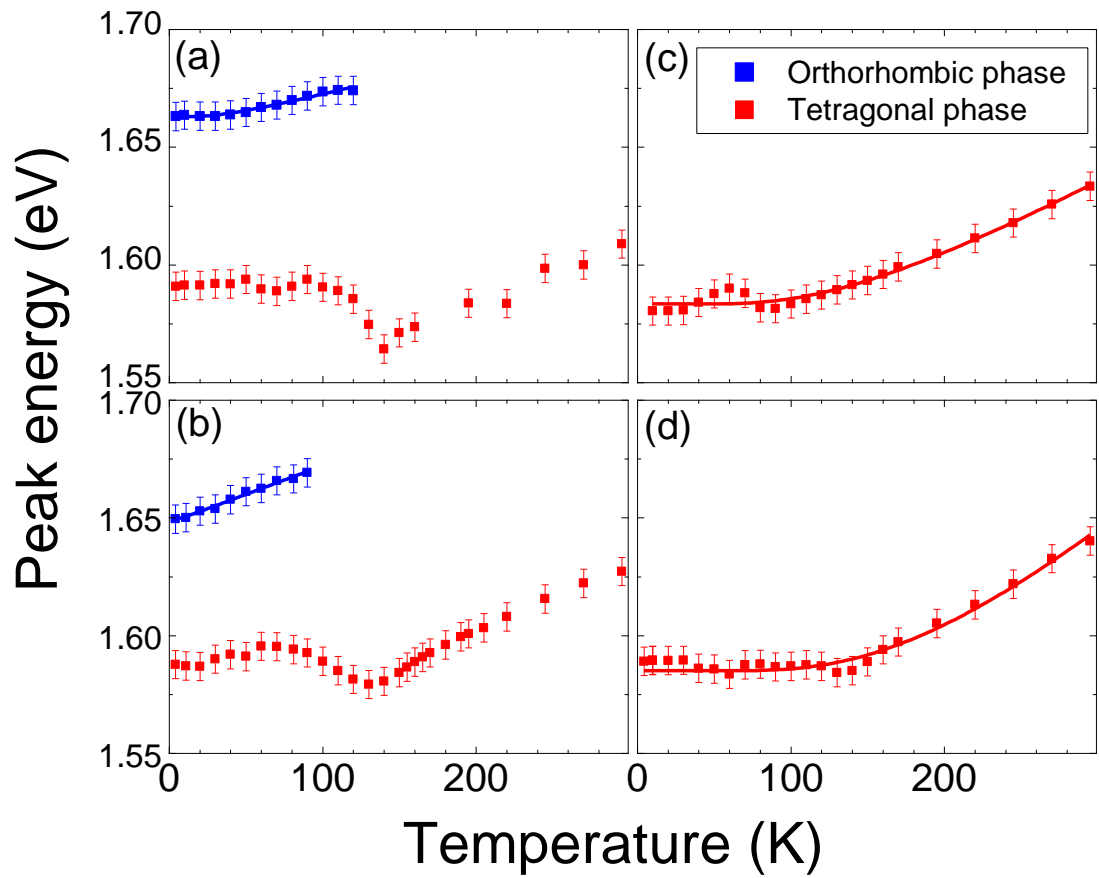

**Figure 6**

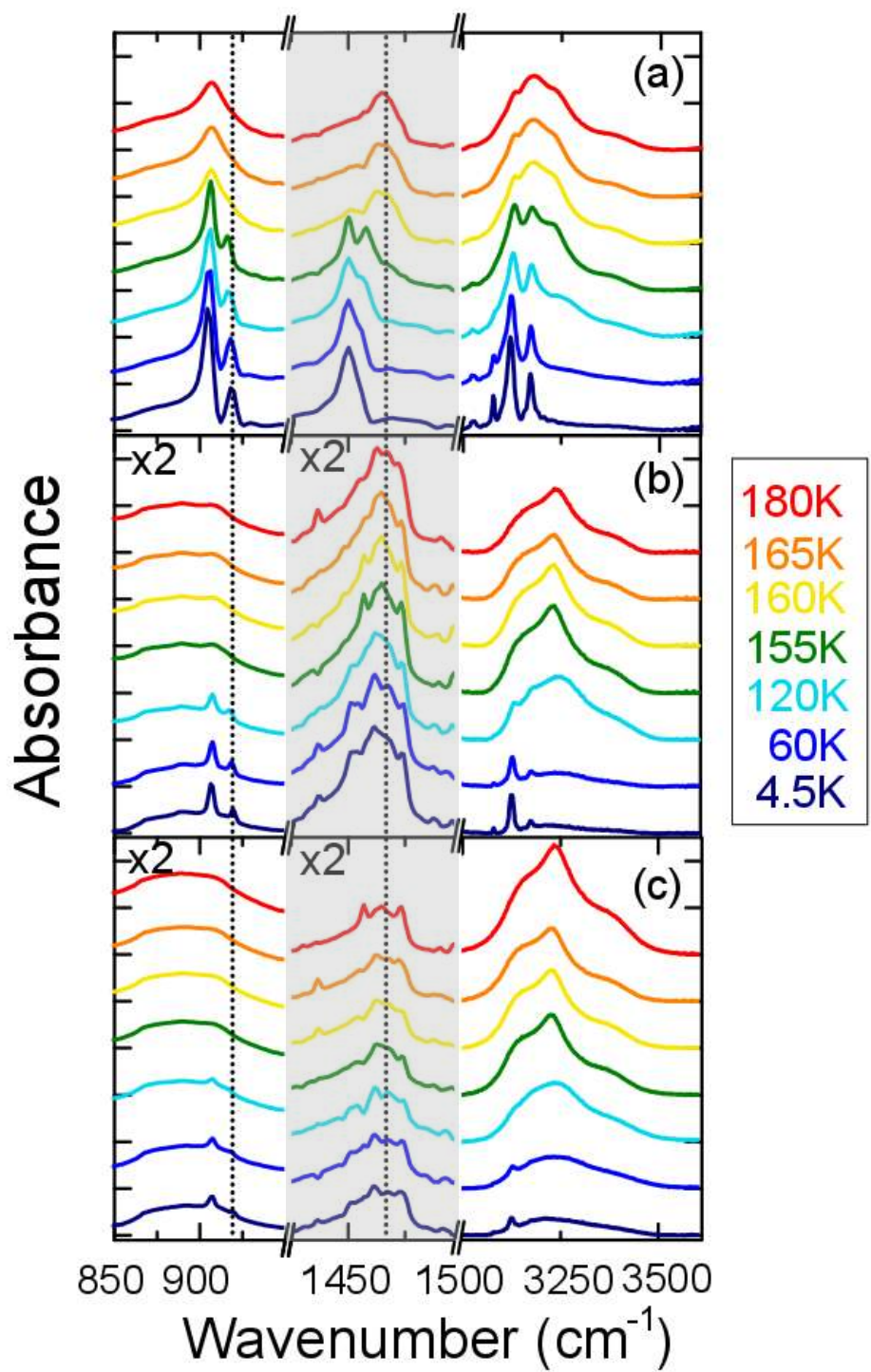

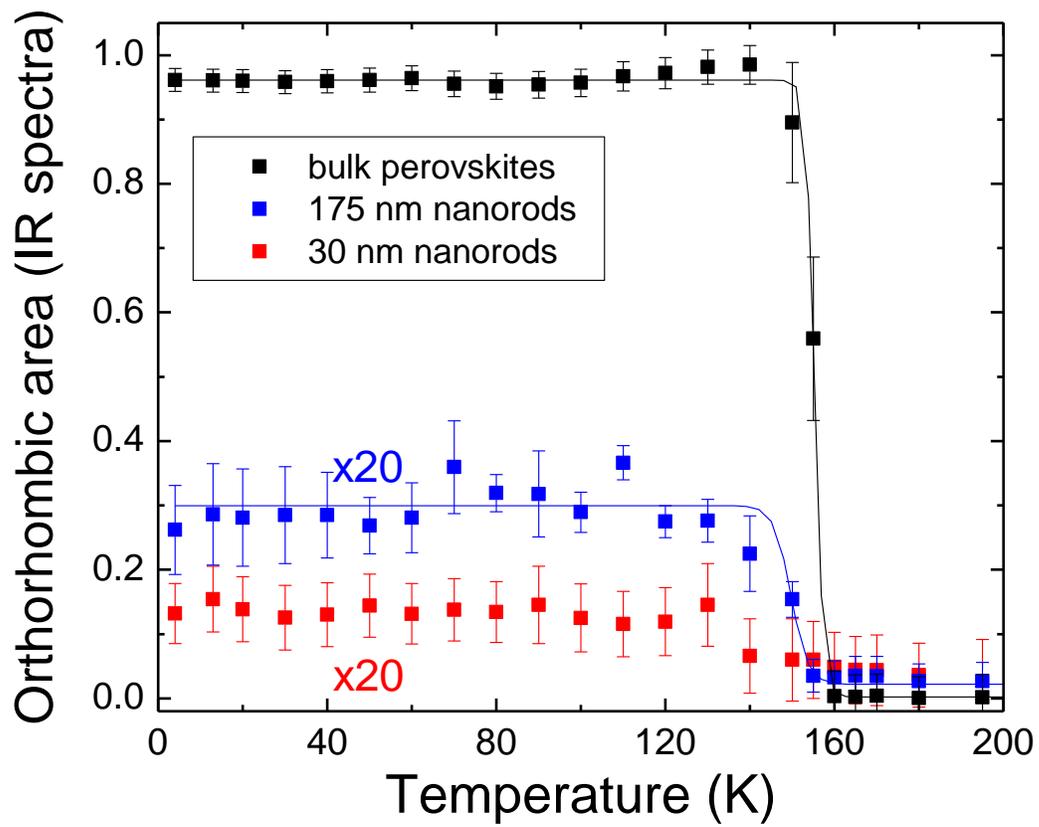

**Figure 8**

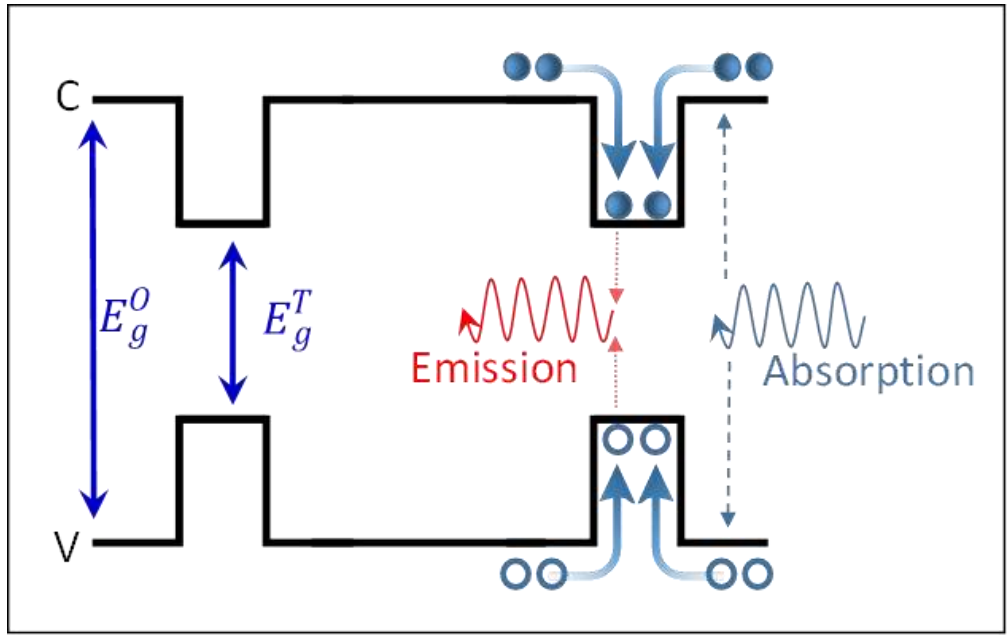

**Figure 9**